\documentclass[twocolumn,superscriptaddress,amsmath,amssymb,aps,pra]{revtex4}
\usepackage{amsmath,amsthm,amssymb}
\usepackage{latexsym}
\usepackage{amscd,graphicx}
\usepackage[titletoc]{appendix}
\usepackage{epsfig}
\usepackage{epstopdf}
\usepackage[normalem]{ulem}
\usepackage[dvipsnames]{xcolor}
\usepackage[colorlinks=true,linkcolor=red,citecolor=blue]{hyperref}

\begin{document}

\title{Cavity magnon–polariton interface for strong	spin–spin coupling}

\author{Ma-Lei Peng}
\affiliation{Department of Physics, Wenzhou University, Zhejiang 325035, China}

\author{Miao Tian}
\affiliation{Department of Physics, Wenzhou University, Zhejiang 325035, China}

\author{Xue-Chun Chen}
\affiliation{Department of Physics, Wenzhou University, Zhejiang 325035, China}

\author{Ming-Feng Wang}
\affiliation{Department of Physics, Wenzhou University, Zhejiang 325035, China}

\author{Guo-Qiang Zhang}
\affiliation{School of Physics, Hangzhou Normal University, Hangzhou, Zhejiang 311121, China}

\author{Hai-Chao Li}
\affiliation{College of Physics and Electronic Science, Hubei Normal University, Huangshi 435002, China}
\affiliation{Laboratory of Solid State Microstructures, Nanjing University, Nanjing 210093, China}

\author{Wei Xiong}
\altaffiliation{xiongweiphys@wzu.edu.cn}
\affiliation{Department of Physics, Wenzhou University, Zhejiang 325035, China}
\affiliation{Laboratory of Solid State Microstructures, Nanjing University, Nanjing 210093, China}

\date{\today}

\begin{abstract}
Strong coupling between single qubits is crucial for quantum information science and quantum computation. However, it is still challenged, especially for single solid-state qubit. Here, we propose a hybrid quantum system, consisting of a coplanar waveguide (CPW) resonator weakly coupled to a single nitrogen-vacancy spin in diamond and a yttrium-iron-garnet (YIG) nanosphere holding Kerr magnons, to realize strong long-distance spin-spin coupling.  With a strong driving field on magnons, the Kerr effect can squeeze magnons, and {thus the coupling between the CPW resonator and the sequeezed magnons is exponentially enhanced}, which produces two cavity-magnon polaritons, i.e., the high-frequency polariton (HP) and low-frequency polariton (LP). When the enhanced cavity-magnon coupling {approaches} the critical value (i.e., the frequency of the LP becomes zero), the spin is fully decoupled from the HP, while the coupling between the spin and the LP is significantly improved. In the dispersive regime, a strong spin-spin coupling mediated by the LP is achieved with accessible parameters. Our proposal indicates that the critical cavity-magnon polarition is a potential interface to realize strong spin-spin coupling and manipulates remote solid spins.
\end{abstract}


\maketitle

Solid spins such as nitrogen-vacancy centers in diamond~\cite{Schirhagl-2014}, having good tunability~\cite{doherty2013} and long coherence time~\cite{Gill2013}, are regarded as promising platforms for quantum information science.  However, direct spin-spin coupling is weak due to their small magnetic dipole moments~\cite{zhu2011,Xiong-2021}. To overcome this, the natural ideal is to look for quantum interfaces~{\cite{Li-2016,Rabl-2010,Xiong-2021,Hei-2023}} as bridges to couple spins, forming diverse hybrid quantum systems~\cite{xiang2013,kurizki2015}. Recently, the emerged low-loss magnons (i.e., the quanta of collective spin excitations) in ferromagnetic materials~\cite{Yuan-2021} have shown great potential in achieving strong spin-spin coupling~\cite{neuman-2020}. For example, linear~\cite{neuman-2020} and nonlinear~\cite{Xiong2-2022} magnons in yttrium-iron-garnet (YIG) nanospheres have been proposed to realize strong spin-spin coupling. In addition, strong spin-photon coupling in a microwave cavity can also be demonstrated by the YIG nanosphere~\cite{hei-2021}. Besides these, magnons in a bulk material~\cite{trifunovic-2013,Fukami-2021} and a thin ferromagnet film~\cite{Skogvoll-2021} have been suggested to coherently couple remote spins. However, achieving strong spin-spin coupling is still a challenge.

In this letter, we propose a novel approach to realize strong spin-spin coupling in a hybrid cavity-magnon-spin system, where the spin in diamond and Kerr magnons (i.e., magnons with Kerr effect) in a YIG nanosphere are weakly coupled to the photons in a coplanar waveguide (CPW) resonator. Experimentally,  the strong and tunable magnon Kerr effect, {originating from the magnetocrystalline anisotropy~\cite{zhang-2019}, has been demonstrated~\cite{wang-2018}} and gives rise to rich phenomena~\cite{wang-2018,Shen-2022,JChen-2023,Liu-2023}. Under a strong driving field, the Kerr effect can {\it squeeze} magnons, and thus the coupling between magnons and the CPW resonator is exponentially enhanced to the strong coupling regime. The strong magnon-photon coupling generates two polaritons, i.e., the high-frequency polariton (HP) and the low-frequency polariton (LP). We show that the LP exhibits criticality when the enhanced magnon-photon coupling {approaches} the critical point. Around this point, the spin-LP coupling is {\it greatly enhanced} while the spin-HP coupling is {\it fully suppressed}. {Considering that two separated spins are dispersively coupled to the LP}, a tunable and strong spin-spin coupling can be induced by adiabatically eliminating the LP, as demonstrated by the Rabi oscillation between two spins. Our proposal indicates that the critical cavity-magnon polariton is a promising interface to realize strong spin-spin coupling and manipulates solid-state qubits.

\begin{figure}
	\includegraphics[scale=0.38]{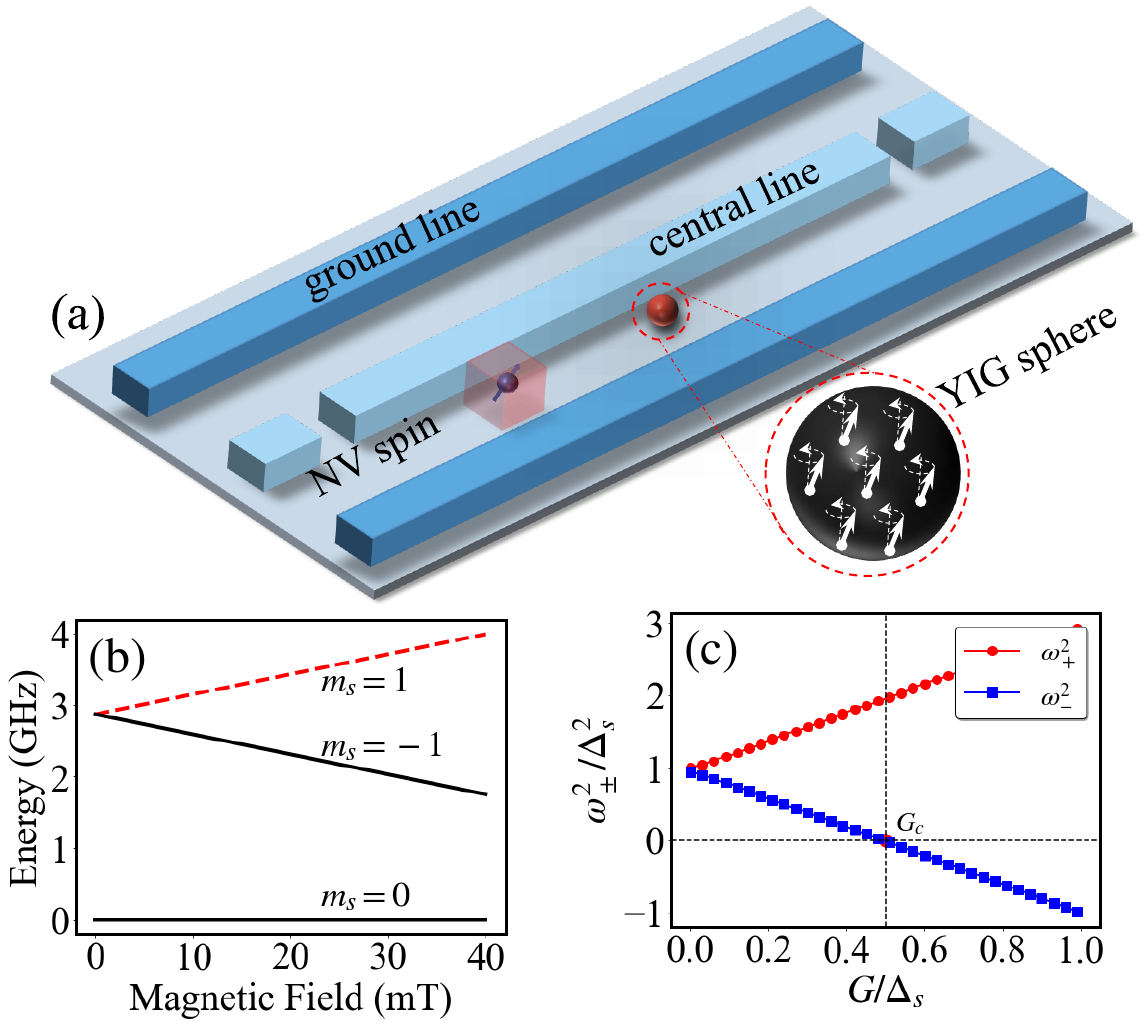}
	\caption{(a) Schematic diagram of a hybrid quantum system. The single nitrogen vacancy (NV) center spin {(pink semi-transparent boxes with a dot)},  located near the central line {(light blue strip)}, is weakly coupled to the CPW resonator {(the central part of the central line)}. {The dark blue strip denotes the ground line of the CPW resonator}. (b) The level structure of the triplet ground state of the NV center, $m_s=0$ and $m_s=-1$ are selected to form a spin qubit. (c) The square of the normalized polariton frequencies versus the normalized coupling strength between the squeezed magnons and the CPW resonator. $G_c$ is the critical coupling strength.} \label{fig1}		
\end{figure}

We consider a hybrid system consisting of a CPW resonator weakly coupled to both a single NV spin in diamond and Kerr magnons in the Kittel mode of a YIG nanosphere, as shown in Fig.~\ref{fig1}(a). The Kittel mode is driven by an external field with Rabi frequency $\Omega_d$ and frequency $\omega_d$. In the rotating frame with respect to the driving field, the Hamiltonian of the system is (setting $\hbar=1$)
\begin{align} \label{eq1}
	H_{\rm sys}=&\frac{1}{2} \Delta_{\rm NV} \sigma_z+\Delta_a a^\dag a+g_m\left(a^{\dagger} m+a m^{\dagger}\right)\nonumber\\
	&+\lambda\left(\sigma_{+}a+a^{\dagger} \sigma_{-}\right)+H_K+\Omega_d\left( m^{\dagger}+m\right),
\end{align}
where $\Delta_{\rm NV}=\omega_{\rm NV}-\omega_d$ with $\omega_{\rm NV}$ being the transition frequency between the lowest two levels of the triplet ground state of the NV~[see Fig.~\ref{fig1}(b)], $\Delta_a=\omega_c-\omega_d$ with $\omega_c$ being the frequency of the CPW resonator, $g_m$ is the magnon-photon coupling strength~\cite{hei-2021}. For a nanosphere with the radius $R\sim50$ nm, $g_m/2\pi\sim0.2$ MHz, which is much smaller than the typical decay rates of the CPW resonator~{($\kappa_c/2\pi\sim 1$ MHz)~\cite{Niemczyk-2009} and the Kittel mode~($\kappa_m/2\pi\sim 1$ MHz)~\cite{Tabuchi-2014}}, i.e., $g_m<\kappa_c,\kappa_m$. Thus, the magnon-photon coupling is in the weak coupling regime for the nanosphere. $\lambda$ is the spin-photon coupling strength. For $\omega_c/2\pi\sim6.78$ GHz, the inductance of the CPW resonator $L\sim2$ nH, the distance between the spin and the the central line of the CWP resonator $d\sim50$ nm, and the magnetic field generated by the vacuum fluctuation of photons $B_{0,f}\sim2.5\times10^{-7}$ T, $\lambda\sim7$ kHz is estimated~\cite{Twamley2010}, which is much smaller than $\kappa_c$. $H_K$ in Eq.~(\ref{eq1}) denotes the Hamiltonian of the Kerr magnons, given by~\cite{wang-2018}
$H_K=\delta_m m^\dag m+K m^{\dagger} m^{\dagger} m m,$
where $\delta_m=\omega_m-\omega_d$ with  the frequency of the {Kittel} mode $\omega_m$ propotional to the biased magnetic field $B_0$, and $K=\mu_{0}K_{\rm an}\gamma^2/M^{2}V_{m}$ is the coefficient of the Kerr nonlinearity, with the gyromagnetic ration {$\gamma/2\pi=g_e\mu_B/\hbar=28$ GHz/T}, the vacuum permeability $\mu_0$, the first-order anisotropy constant of the YIG sphere $K_{\rm an}$, the saturation magnetization $M$, and the volume of the YIG nanosphere $V_m$.  Experimentally, the biased magnetic
field $B_0$ can be generated by a superconducting magnet to magnetize the YIG sphere, and it is tunable in the range of $0$ to $1$ T, so the given frequency of the magnons in the Kittel mode ranges from several hundreds of megahertz to $28$ GHz~\cite{wang-2018}. {We here take the experimentally accessible value $B_0=98.5$ mT for achieving $\omega_m/2\pi=2.6$ GHz}. Other parameters are $\mu_0 K_{\rm an}=2480~{\rm J/m^3}$, $M=196$ kA/m~\cite{zhang-2019}. Apparently, the Kerr coefficient is inversely proportional
to the volume of the YIG sphere, i.e., $K\propto1/V_m$. This indicates that the Kerr effect becomes significant for nanospheres. For example, when $R\sim50$ nm, $K/2\pi\sim128$ Hz, but $K/2\pi\sim0.05$ nHz for $R\sim0.5$ mm (the usual size of the YIG sphere used in current experiments). Obviously, $K$ is much smaller in the latter case. Because our proposal mainly relies on the Kerr effect,  we here use the YIG nanosphere to obtain a strong Kerr effect.

For a strong driving field, the Hamiltonian $H_{\rm sys}$ in Eq.~(\ref{eq1}) can be linearized~\cite{Xiong2-2022} as
\begin{align}
	H_{\text {lin}}= &\frac{1}{2} \Delta_{\rm NV} \sigma_z+\Delta_a a^\dag a+\Delta_m m^{\dagger} m+K_s\left(m^2+m^{{\dagger}, 2}\right)\nonumber\\
	&+\lambda\left(\sigma_{+}a+a^{\dagger} \sigma_{-}\right)+g_m\left(a^{\dagger} m+a m^{\dagger}\right), \label{eq:2}
\end{align}
where the effective magnon {frequency} detuning $\Delta_m=\delta_m+4 K|\langle m\rangle|^2$ is induced by the Kerr effect, which has been demonstrated experimentally~\cite{wang-2018}. The amplified coefficient $K_s=K\langle m\rangle^2$ is the effective strength of the two-magnon process, which can give rise to squeeze magnons in the {Kittel} mode. Aligning the biased magnetic field along the crystalline axis $[100]$ {($[110]$)} of the YIG sphere~\cite{zhang-2019,wang-2018}, $K$ is positive {(negative)}, and {thus we have $K_s > 0$ ($ K_s < 0$)}. {To achieve our goal, we take $K_s<0$ below}. By further applying a Bogoliubov transformation $m=m_s\cosh\left(r_m\right)+m_s^{\dagger}\sinh\left(r_m\right)$ with the squeezing parameter $ r_m=\frac{1}{4} \ln \frac{\Delta_m-2 K_s}{\Delta_m+2 K_s}$, Eq.~(\ref{eq:2}) becomes
$H_{\rm SQ}=\frac{1}{2} \Delta_{\rm NV} \sigma_z+H_{\rm CMS}+\lambda\left(\sigma_{+} a+a^{\dagger} \sigma_{-}\right)$, with
\begin{align}\label{eq2}
	H_{\rm CMS}=\Delta_a a^{\dagger} a+\Delta_s m_s^{\dagger} m_s+G\left(a^{\dagger}+a\right)\left(m_s^{\dagger}+m_s\right)
\end{align}
being the effective Hamiltonian of the CPW resonator coupled to the squeezed magnons with the effective frequency $\Delta_s=\sqrt{\Delta_m^2-4 K_s^2}$, $G=\frac{1}{2}g_m e^{r_m}$ is the exponentially enhanced coupling strength between the squeezed magnons and the CPW resonator. When $r_m=3$ ($5$), $G/2\pi=2$ ($17$) MHz for $R=50$ nm. This shows that strong coupling between the squeezed magnons and the CPW resonator can be realized by tuning the squeezing parameter $r_m$. In addition, $G$ can be further enhanced by using the larger radius of the YIG sphere when $r_m$ is fixed. Once the strong coupling between the squeezed magnons and the CPW resonator is achieved, the {counter-rotating} terms $\propto a^{\dagger}m_s^{\dagger} $ and $am_s$ in Eq.~(\ref{eq2}) are related to two-mode squeezing, while rotating terms $\propto a^\dag m_s$ and $a m_s^\dag$ allow quantum state transfer between the squeezed magnons and the CPW resonator. By combining these, HP and LP can be given by further diagonalizing the Hamiltonian $H_{\rm CMS}$ in Eq.~(\ref{eq2}) as $H_{\rm diag}=\omega_{+} a_{+}^{\dagger} a_{+}+\omega_{-} a_{-}^{\dagger} a_{-}$, where the corresponding eigenfrequencies are
\begin{equation}
	\omega_{\pm}^2=\frac{1}{2}\left[\Delta_a^2+\Delta_s^2 \pm \sqrt{\left(\Delta_a^2-\Delta_s^2\right)^2+16 G^2 \Delta_a \Delta_s}\right].\label{eq:5}
\end{equation}
In the polariton representation, $H_{\rm SQ}$ can be expressed as
\begin{align}\label{eq:8}
	H_{\rm CMP}=&\frac{1}{2} \Delta_{\rm NV} \sigma_z+\omega_{+} a_{+}^{\dagger} a_{+}+\omega_{-} a_{-}^{\dagger} a_{-}\\
	&+{g_+}\left(\sigma_{+} a_{-}+\sigma_{-} a_{-}^{\dagger}\right)+{g_-}\left(\sigma_{+} a_{-}^{\dagger}+\sigma_{-} a_{-}\right)\nonumber\\
	&+{g_+^\prime}\left(\sigma_{+} a_{+}+\sigma_{-} a_{+}^{\dagger}\right)+{g_-^\prime}\left(\sigma_{+} a_{+}^{\dagger}+\sigma_{-} a_{+}\right),\notag
\end{align}
where $g_{\pm}=\lambda\cos\theta\left(\Delta_a\pm\omega_{-}\right)/2\sqrt{\Delta_a\omega_{-}}$ denote the effective coupling strength between the NV spin and the LP, $g^\prime_{\pm}=\lambda\sin\theta\left(\Delta_a\pm\omega_{+}\right)/2\sqrt{\Delta_a\omega_{+}}$ represent the effective coupling strength between the NV spin and the HP. Obviously, both $g_{\pm}$ and $g^\prime_{\pm}$ can be {\it tuned} by the driving field on the {Kittel} mode of the YIG nanosphere. The parameter $\theta$ is defined by $\tan(2\theta)=4G\sqrt{\Delta_a \Delta_s}/(\Delta_a^2-\Delta_s^2)$. To show the behavior of two polaritons with the coupling strength $G$, we plot the square of polariton frequencies versus the coupling strength $G$ in Fig.~\ref{fig1}(c). Clearly, one can see that $\omega_{+}^2$ increases with $G$, but $\omega_{-}^2$ decreases. When $\omega_-^2=0$, $G$ approaches the critical value $G_c\equiv\frac{1}{2}\sqrt{\Delta_a\Delta_s}$, which means $\omega_-$ is real for $G<G_c$, while $\omega_-$ is imaginary for $G>G_c$. When we operate the cavity-magnon subsystem around the critical point (i.e., $G\rightarrow G_c$) and $\Delta_a\gg\Delta_s$ is satisfied, we have ${g_+}\approx {g_-}\rightarrow \frac{{1}}{2}\lambda\sqrt{\Delta_a/\omega_{-}}$, ${g_+^\prime}\approx {g_-^\prime}\rightarrow0$. Due to the large $\Delta_a$ and the extremely small $\omega_-$, ${g_+}\approx {g_-}\gg\lambda$.  These indicate that the spin-HP coupling can be completely {\rm suppressed}, while the spin-LP coupling is {\rm significantly enhanced}. {By choosing $\Delta_a=4\times10^3\omega_{-}$ with $\omega_-/2\pi=1.6$ MHz, ${g_+}={g_-}\sim 31.6\lambda$ are obtained}. Using $d=50$ nm, $\lambda=2\pi\times7$ kHz, resulting in {${g_+/2\pi}\sim 0.22$} MHz. This suggests that the coupling between the spin and the LP can be in the strong coupling regime. In principle, ${g_{\pm}}$ can be further enhanced by using the larger $\Delta_a$ or much smaller $\omega_-$.

\begin{figure}
	\includegraphics[scale=0.6]{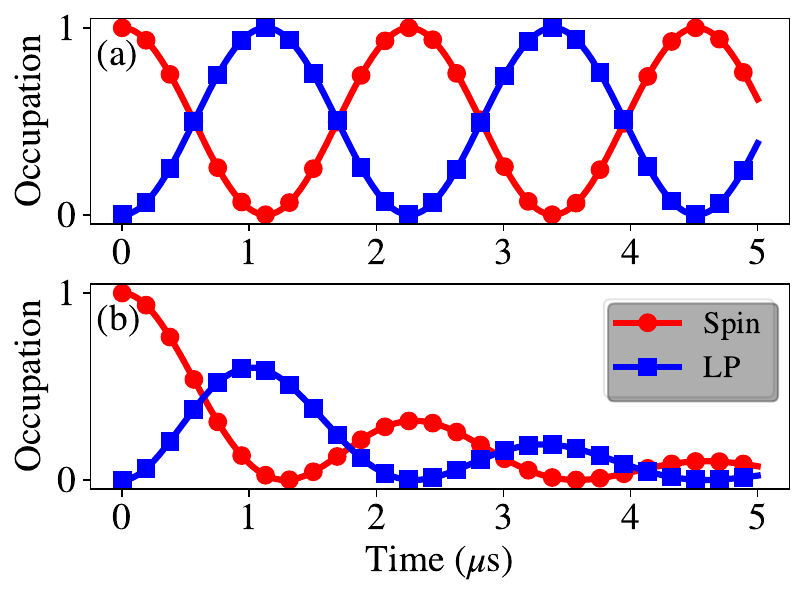}
	\caption{The occupation of the LP and spin qubit versus the evolution time at $G\rightarrow G_c$ and $\Delta_a\gg\Delta_s$ (a) without and (b) with dissipations. The spin decay rate is $\gamma_{\perp}/2\pi\sim1$ kHz and the LP {decay} rate is $\kappa_-/2\pi\sim1$ MHz. In both (a) and (b), the spin qubit is initially prepared in the excited state and the LP is in the ground state, and the coupling strength is $g_+/2\pi=0.22$ MHz.}\label{fig2}
\end{figure}
\begin{figure}
	\includegraphics[scale=0.6]{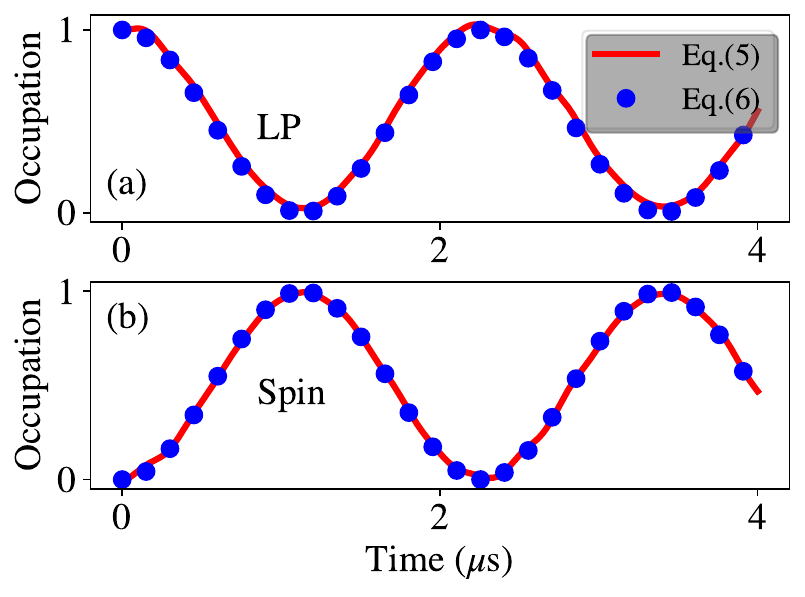}
	\caption{{The occupation of the LP and spin qubit versus the evolution time at $G\rightarrow G_c$ and $\Delta_a\gg\Delta_s$ (a) without and (b) with dissipation, where the spin (LP) is initially prepared in the excited (ground) state. Other parameters are the same as those in Fig.~{\ref{fig2}}}.}\label{fig3}
\end{figure}

{When the condition $\Delta_{\rm NV}\geq\omega_{-}\gg g_\pm$ is ensured, the rotating-wave approximation can be safely applied to Eq.~(\ref{eq:8}){, as numerically demonstrated in Fig.~\ref{fig3}}. To satisfy this condition, the frequency of the driving field} $\omega_d/2\pi=2.598$ GHz and $\omega_{\rm NV}/2\pi=2.6$ GHz are taken. Neglecting the counter-rotating terms, Eq.~(\ref{eq:8}) reduces to the { Jaynes-Cummings} model
\begin{equation}
	H_{\rm JC}=\frac{1}{2} \Delta_{\rm NV}\sigma_z+\omega_{-} a_{-}^{\dagger} a_{-}+{g_+}\left(\sigma_{+} a_{-}+\sigma_{-} a_{-}^{\dagger}\right).
	\label{eq:10} 
\end{equation}
This Hamiltonian allows quantum state exchange between the spin and the LP, as demonstrated in Fig.~\ref{fig2}(a), where the spin is initially  prepared in the excited state and the LP is in the ground state.

When dissipations are included, the dynamics of the system can be described by the master equation,
\begin{equation}
	\frac{d \rho}{d t}=-i\left[H_{\mathrm{JC}}, \rho\right]+\kappa_- D\left[a_{-}\right] \rho+\gamma_{\perp} D\left[\sigma_{-}\right] \rho,
	\label{eq:11}
\end{equation}
where $D[o] \rho=o \rho o^{\dagger}-\frac{1}{2}\left(o^{\dagger} o \rho+\rho o^{\dagger} o\right)$, and $\gamma_{\perp}$ is the transversal (longitudinal) relaxation rate of the NV spin~\cite{Angerer17}, $\kappa_-$  is the decay rate of the LP. In Fig.~\ref{fig2}(b), we use the qutip package in python \cite{Johansson1,Johansson2} to numerically simulate the dynamics of the spin and LP governed by Eq.~(\ref{eq:11}). The results show that state exchange between the spin and the LP can be realized in the presence of dissipation such as $\kappa_-/2\pi\sim 1$ MHz and $\gamma_{\perp}/2\pi\sim 1$ kHz~\cite{LiB-2019}.
\begin{figure}
	\includegraphics[scale=0.6]{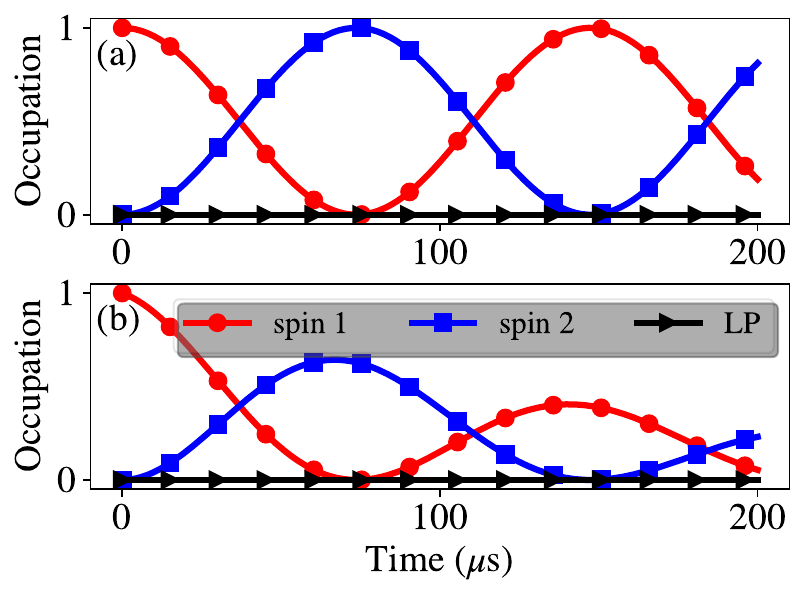}
	\caption{The occupation of two spins and the LP versus the evolution time in the dispersive regime (a) without and (b) with dissipation  are simulated using the effective Hamiltonian in Eq.~(\ref{eq12}). The parameters $\omega_-/2\pi=1.6$ MHz, $\Delta_{\rm NV}/2\pi=16$ MHz, and $g_+/2\pi=0.22$ MHz are used.}\label{fig4}
\end{figure}

Here, we further consider the case that two identical NV spins are symmetrically placed away from the YIG sphere in the CPW resonator. Thus, two spins interact with the CPW resonator with the same coupling strength $\lambda$.  By operating the cavity-magnon subsystem around the critical point, the couplings between two spins and the HP can be fully suppressed, while the couplings between two spins and the LP {are} greatly enhanced, similar to the single spin case. Therefore, the Hamiltonian of the hybrid system with two identical spins can be effectively described by {Tavis-Cummings} model,
\begin{align}
	H_{\rm TC}=&\omega_{-} a_{-}^{\dagger} a_{-}+\sum_{j=1,2}[\frac{\Delta_{\rm NV}}{2}\sigma_z^{(j)}
	+{g_+}(\sigma_{+}^{(j)} a_{-}+{\rm H.c.})].\label{eq12}
\end{align}
{To enter the dispersive regime, i.e., $|\Delta_{\rm NV}-\omega_-|\gg {g_+}$, {$\Delta_{\rm NV}/2\pi=16$ MHz is chosen.} This can be realized by tuning the magnetic field acting on the transition between levels $m_s=\pm1$. {In this regime}, the LP can be as an interface to induce an indirect and tunable coupling between two spins by using the Fr\"{o}hlich-Nakajima transformation~\cite{Frohlich,Nakajima}. Up to the second order in ${g_+}/\Delta_{\rm NV}$, the spin qubits are decoupled from the LP. By adiabatically eliminating the LP, the effective spin-spin coupling Hamiltonian is 
	\begin{equation}
		H_{\text {eff}}=\frac{1}{2} \omega_{\rm eff} \left(\sigma_z^{(1)}+\sigma_z^{(2)}\right)+g_{\text {eff}}\left(\sigma_{+}^{(1)} \sigma_{-}^{(2)}+{\rm H.c.}\right),
		\label{eq:12} 
	\end{equation}
	where $\omega_{\rm eff}=\Delta_{\rm NV}+2g_{\rm eff}n_-+g_{\rm eff}$ is the effective transition frequency of the NV spin, depending on the mean occupation number $n_-=\langle a_{-}^{\dagger} a_{-}\rangle$ of the LP, $g_{\text {eff}}=-{g_+^2}/(\Delta_{\rm NV}-\omega_-)$ is the effective spin-spin coupling strength induced by the LP. To estimate $g_{\rm eff}$, we assume the distance between the spin and the central line of the CPW resonator $d=50$ nm, so {${g_+/2\pi}=0.22$ MHz},  thus we have {$g_{\rm eff}/2\pi=3.4$ kHz at $\Delta_{\rm NV}/2\pi=16$ MHz}. Obviously, $g_{\text {eff}}>\gamma_{\perp}~(\sim1$ kHz), i.e., the strong spin-spin coupling is achieved. This can be directly demonstrated by simulating the dynamics of the original system [see  Eq.~(\ref{eq12})] with the master equation in the dispersive regime. In Figs.~\ref{fig4}(a) and \ref{fig4}(b), quantum state exchange (i.e., Rabi oscillation) between two separated spins can be clearly observed in both the absence and presence of dissipation, while the LP is always in the initial state. {Note that the simulation result in Fig.~\ref{fig4}(a) can also be obtained by directly solving the Schrodinger equation with the Hamiltonian in Eq.~(\ref{eq:12}). Specifically, the probability of the spin 1 in the excited state is $P_{\rm e}^{(1)}=\cos^2 (g_{\rm eff}t)=\langle \sigma_z^{(1)}\rangle$ and the probability of the spin 2 in the excited state is $P_{\rm e}^{(2)}=\sin^2 (g_{\rm eff}t)=\langle \sigma_z^{(2)}\rangle$. When dissipation is included, the simulation result in Fig.~\ref{fig4}(b) can be given by $P_{\rm e}^{(1)}=\exp(-\gamma_\perp t)\cos^2 (g_{\rm eff}t)=\langle \sigma_z^{(1)}\rangle$ and $P_{\rm e}^{(2)}=\exp(-\gamma_\perp t)\sin^2 (g_{\rm eff}t)=\langle \sigma_z^{(2)}\rangle$. This analytical expression can be obtained by solving the condition Hamiltonian $H_{\rm con}=H_{\rm eff}-\frac{i}{2}\gamma_\perp[\sigma_z^{(1)}+\sigma_z^{(2)}$].}


	In summary, we have proposed a hybrid system consisting of a CPW resonator weakly coupled to NV spins and a YIG nanosphere supporting magnons with {the} Kerr effect. With a strong driving field, the Kerr effect can squeeze magnons, giving rise to an exponentially enhanced strong cavity-magnon coupling, and thus HP and LP can be formed. {When the cavity-magnon coupling strength reaches the critical value}, the spin-LP coupling is greatly enhanced to the strong coupling regime with accessible parameters, while the coupling between the spin and the HP is fully suppressed. Using the LP as quantum interface in the dispersive regime, {the} strong spin-spin coupling can be achieved, which allows quantum state exchange between two spins. Our scheme provides a potential path to realize strong spin-spin coupling with critical cavity-magnon polaritons.
	\medskip
	
	\noindent{\bf Funding.} Natural Science Foundation of Zhejiang Province (Grant No. LY24A040004), "Pioneer" and "Leading Goose" R\&D Program of Zhejiang (Grant No. 2025C01028), and Shenzhen International Quantum Academy (Grant No. SIQA2024KFKT010). 
	
	\medskip
	\noindent{\bf Disclosures.}  The authors declare no conflicts of interest.
	
	\medskip
	\noindent{\bf Data availability.} Data underlying the results presented in this paper are not publicly available at this time but may be btained from the authors on reasonable request.

\end{document}